\documentclass[a4paper,10pt]{article}
\usepackage{graphicx}
\usepackage{dcolumn}
\usepackage{bm}
\usepackage{amsmath}
\usepackage{amssymb}

\title{Antiferromagnetic Ising spin glass competing with BCS pairing interaction in a transverse field}

\author{S.G. Magalhaes\footnote{ggarcia@ccne.ufsm.br} $^{,a}$, F.M. Zimmer$^a$, C.J. Kipper$^a$, E.J. Callegari$^a$%
\\
\\
{$^a$\normalsize{\it Laborat\'orio de Mec\^anica Estat\'\i stica e Teoria da Mat\'eria Condensada,}}
\\
{\normalsize\it Dep. F\'\i sica, UFSM, 97105-900 Santa Maria, RS, Brazil}
}

\date{}
\begin{document}
\maketitle

\begin{abstract}
The competition among spin glass (SG), antiferromagnetism (AF) 
and local pairing superconductivity (PAIR) 
is studied in a 
two-sublattice fermionic Ising spin glass model with a local BCS pairing interaction 
in the presence of an applied magnetic transverse field $\Gamma$. 
In the present approach, spins in different sublattices interact with a 
Gaussian random coupling with an antiferromagnetic mean $J_0$ and standard deviation $J$. 
The problem is formulated in the 
path integral formalism in which spin operators are represented by 
bilinear combinations of Grassmann variables. 
The saddle-point Grand Canonical potential 
is obtained  
within the static approximation and 
the  replica symmetric ansatz. 
The results are analysed in phase diagrams in which the AF and the SG phases can occur for small $g$ ($g$ is the strength of the local superconductor coupling written in 
units of $J$), while
the PAIR phase appears as unique solution for large $g$.
However, there is a complex line transition 
separating the PAIR phase from the others. It is second order at high temperature that ends in a tricritical point.  
The quantum fluctuations   
affect deeply the transition lines 
and the tricritical point due to the presence of $\Gamma$.  
\end{abstract}
\section{Introduction}
It is now well-established  
that strongly correlated systems such as heavy fermions (HF)
\cite{experimentshv} and high temperature superconductors (HTSC) 
\cite{experimentshtc}, upon doping, can present magnetic order or superconductivity.
The complexity involved in such physical systems as, for instance, 
the existence of Non-Fermi Liquid (NFL) behaviour, has given rise 
to new theoretical approaches.
In particular, recent works have proposed that the  presence of disorder can 
affect these strongly correlated systems (see \cite{Dagotto} and references therein)
being even source of NFL behaviour, for example, in HF \cite{Miranda,Castro}. 
The presence of disorder can also induce frustration which, in fact, has been found in several HF 
\cite{focoexperimentalHF1,focoexperimentalHF2,focoexperimentalHF3,focoexperimentalHF4} 
and HTSC \cite{focoexperimentalHTc1,focoexperimentalHTc2} 
physical systems. Some theories have been investigating whether or not 
spin glass (SG) phase can be found in models designed to study 
certain aspects of HF or  HTSC systems \cite{Galitski,Castro-Neto,Caldeira}. 
For instance, 
the existence of a SG solution has been demonstrated 
in the Kondo lattice model \cite{Theumann,Kiselev}. 
Nevertheless, relatively little consideration \cite{Alvarez} has been given in order to 
obtain the behaviour 
of the transition temperatures for these 
strongly correlated physical systems and, thus, to mimic the  phase boundaries
of their global phase diagram 
which includes antiferromagnetism (AF), SG, 
superconductivity and the possible presence of a Quantum Critical Point  (QCP).        
      
Regarding the experimental scenario, 
there are  some 
similarities in the global phase diagrams between some HF and  
HTSC, although the microscopical mechanisms involved in such systems are 
different \cite{Maple}. 
For instance, the HTSC compound 
$Y_{1-x}Ca_{x}Ba_{2}Cu_{3}O_{6}$ has a phase diagram temperature $T$ {\it versus} 
the hole concentration $p_{sh}$ \cite{focoexperimentalHTc1} which 
displays  an AF ordering for low $p_{sh}$. 
The respective N\'eel temperature $T_{N}$ decreases for  $0<p_{sh}<0.035$ until 
 the onset
of a second 
transition at $T_{f}$ in which 
is found
the superposition of frustration with 
a preformed AF background.
For $p_{sh}>0.035$, there is another transition 
from the previous mixed state to a pure SG.
The SG transition temperature  $T_{g}$ also decreases 
with 
the increase of $p_{sh}$ until the onset of superconductivity (SC). However, 
the SG regime is still found into the SC region. For instance, for
$0.06<p_{sh}<0.10$ there is still traces of spin freezing.
Finally, for large values of $p_{sh}$, the superconductivity 
is dominant. 
From the side of the HF 
systems, the example is the compound $U_{1-x}La_{x}Pd_{2}Al_{3}$. When the doping of $La$ is increased,   
the corresponding phase diagram shows the AF ordering to be replaced by a SG state in the region 
$0.25\leq x\leq 0.65 $. 
At $x=1$, the system is a superconductor. The N\'eel temperature $T_{N}$ is sharply decreased from 
$T=14.6 K$ until 
$T=2.6 K$
when $x$ increases, while the subsequent SG transition temperature $T_{g}$  drops 
to a QCP at $x\approx 0.8$.  In the intermediated doping region 
between the QCP and $x=1$,  a NFL behaviour is observed.

In the last years, several works have been studying the competition between SG and  pairing formation 
in a formulation where the spins are represented as bilinear combination of fermionic creation and 
destruction operators.  
\cite{magal99,magal00,Oppermann1}. 
The model used  in such approaches is 
composed of a random Gaussian coupling between localized spins 
together with
pairing interaction in the real space. 
In fact,  it can be shown that both terms of this simple model have the same origin. They can be derived by eliminating the
conduction electrons, in second order of 
perturbation (see Appendix in Ref. \cite{magal99}), 
from an earlier model introduced to treat conventional superconductor doped with magnetic impurities 
\cite{Nass}.   
A saddle-point solution for the respective grand-canonical potential has been obtained within functional 
integral formalism for the Ising \cite{magal99} and the Heisenberg \cite{magal00} version of the 
model in the half-filled situation using
the static approximation \cite{Bray} and the replica 
symmetry ansatz. For the Ising case, the phase diagram temperature {\it versus} the strength of 
the pairing coupling $g$ 
(in units of $J$ that is the variance of the Gaussian random distributed spin-spin coupling )
displays two phase boundaries. For lower $g$, it is found
a second order line transition 
between paramagnetism (NP) and SG at $T_{g}=0.95J$. For large $g$, 
there is a complex line transition 
$T_{1}(g)$ separating the PAIR phase (where pair formation is found) from the SG and NP phases. 
The line transition $T_{1}(g)$ is a second order for high temperature 
when the boundary is between NP and the PAIR phase. However, 
it becomes first order at lower temperature. 
The corresponding tricritical point $T_{trict}$ 
has been 
located on $T_{1}(g)$ above of $T_{g}$. 
As consequence, the boundary  between SG and the PAIR phase is entirely first order. 
Weak hopping corrections done elsewhere \cite{Oppermann1} demonstrated that the    
phase boundaries described previously with no hopping are essentially maintained.

The model in Ref. \cite{magal99} has 
two important shortcomings. The first one is the lack of quantum spin flipping mechanism which 
would  be able to suppress the transition temperatures leading them to
a QCP. 
Even the Heisenberg 
extension of the problem \cite{magal00}  has been unable to produce a QCP. 
This particular weakness has been corrected  in Ref. \cite{Magal05} by the addition of a transverse field $\Gamma$ 
in the Ising version of the model. The presence of $\Gamma$ has changed 
the behaviour of both transition temperatures $T_{g}$ and $T_{1}(g)$. As long $\Gamma$ 
increases,
the first one moves downwards in the direction of a QCP while 
the second one 
is displaced. 
Therefore, $\Gamma$ has also 
suppressed the PAIR solution 
in the sense that it is necessary larger values of the pairing coupling strength $g$ 
to find the PAIR solution \cite{Magal05}. The tricritical point $T_{trict}$ is 
affected  by the presence of $\Gamma$. The transverse field moves up $T_{trict}$ 
which enlarges the first order portion of $T_{1}(g)$.  
Finally, it has been proposed a relationship 
between $\Gamma$ and $g$ ($J$ is kept constant) 
based on the argument that the pairing and 
RKKY  interaction have the same origin in the derivation of the model \cite{magal99}. 
As consequence, the effects described previously 
are superposed in a single phase diagram 
$T$ {\it versus} $g$ due to the increase of $g$,  hence
$\Gamma$.  It shows $T_g$ decreasing towards a QCP at $g=g_c$, then a PAIR phase can be 
found at $g>g_c$ with  $T_{trict}$ located at higher values of $T$ and $g$ than the case $\Gamma=0$.          

Nevertheless, the model used in Refs. \cite{magal99,magal00,Oppermann1} has a second shortcoming, 
it is 
unable to produce an AF solution. Thus, the model is useless if one is trying to study 
the phase boundaries
 which include SG, PAIR phase and also AF.
 However, quite recently the competition 
between AF and SG has been analysed in a disordered two-sublattice fermionic spin model. 
The model is a Gaussian random coupling with an 
antiferromagnetic mean $J_0$  and standard deviation $J$  between
spins in distinct sublattices with
the 
presence of a transverse 
$\Gamma$ and parallel $H$  magnetic fields \cite{Zimmer06}.  
In fact, 
it is the fermionic version 
of the 
model introduced by Korenblit and Shender (KS) \cite{KS} 
 which is used
to study the competition between AF and SG with 
classical Ising variables. This classical two-sublattice model has itself
unexpected effects as compared with the classical single lattice SG Ising model \cite{SK}. 
For example, 
opposite solutions are enforced by the   
degree of frustration $(J_{0})^{-1}$ 
and $H$  (given in units of $J$). 
When degree of frustration decreases 
the AF solution is favored,
while the field $H$ can eventually enhance the frustration  in a certain range \cite{KS}. 
This last effect is related 
with the asymmetry between the two-sublattice due the coupling 
with $H$.  
The presence of the $\Gamma$ in the fermionic version of KS model
has introduced important differences 
if
compared with its classical counterparts 
as long there is a 
competing mechanism associated with $J_{0}$, $H$ and $\Gamma$ 
\cite{Zimmer06}.  
 For instance, $\Gamma$ suppresses the magnetic orders leading their critical temperatures to QCPs, while $H$ 
can favour frustration at the same time that it destroys the AF phase.    

Therefore, the purpose of the present work is to study the competition 
among AF, SG and the PAIR phase using the fermionic Ising KS model with 
a  local pairing interaction
in each sublaticce in the presence of a 
transverse field $\Gamma$. Particularly, the focus is to describe the behaviour 
of the possible transition temperatures present in the problem.   
We follow the same approach used in 
Refs. \cite{magal99,magal00,Magal05}. The partition function is obtained 
in the functional integral formalism where the spin operators 
are given as bilinear combinations of Grassmann fields. The static approximation (SA)
and the replica symmetry (RS) ansatz are used to calculate the saddle-point 
Grand canonical potential. Particularly, we extend for the present two-sublattice problem a procedure 
that mixes  
Nambu matrices and spinors as already introduced in Ref. \cite{Magal05}. 
The stability of the RS solution is also investigated.   
Surely, the simple model used in the present work is not suited to describe 
the extremely complicate physics present in HF as well as in HTSC systems.
However, it can be, at least, useful to  mimic
general features 
of a phase diagram 
in which AF, SG,  pairing 
coupling and a  quantum spin flipping mechanism are present. 

The use of SA and RS ansatz deserves some remarks. 
It is well known the SA is not adequate to describe the low temperature 
behaviour of the spin-spin correlation function \cite{Bray}. However, the  
justification for the use of the SA in this work is based on the fact that our  interest 
is mainly to obtain the possible transition temperatures  associated with AF, SG and PAIR 
competition. 
The analysis of the quantum rotor model  in the $M\rightarrow \infty$ indicates 
that the critical line can be obtained from the zero frequency mode \cite{Sachato}. 

Although the focus in this work is to obtain the phase boundaries 
of the fermionic version of KS model 
with a pairing  interaction in the presence of
a transverse field $\Gamma$,
the thermodynamics is not
the only source of 
valuable information. 
In the case of Ising SG fermionic model,
other physical quantities can have an interesting behaviour. For example,
the density of states (DOS) and, hence, the local Green's function are affected at low temperature 
by the   
replica symmetry breaking \cite{ROppermann}.  In particular, there is a presence 
of a pseudogap.
Recently, 
a mapping between the  one-lattice fermionic Ising SG and the classical 
 Ghatak-Sherrington model \cite{gs}
has demonstrated that
the true origin of such effects in the DOS 
are, 
in fact,  classical \cite{Isaac}. 
However, for the present model we
can speculate if there 
is such kind of mapping due to the presence of the transverse field.  
In that sense, this work can be also thought as a first step towards 
the understanding of the problem given by the model 
introduced here from a many-body perspective 
since the thermodynamics is well understood, 
at least, at mean field level.

This paper presents
the following structure. In section 2, we derive 
the saddle-point Grand Canonical potential and the set of equations for the order parameters 
which is enlarged when compared with a single lattice Ising model studied in Ref. \cite{Magal05}. 
In section 3, we solve 
the order parameter equations. In order to capture properly 
the competition among the phases present in the problem, we build up  
phase diagrams $T$ {\it versus} $g$ 
for 
 several
values of $\Gamma$  and  $J_{0}$  given in units of $J$  where $J_{0}$ and  $J$ are the mean and standard deviation 
of the  random Gaussian spin-spin interlattice 
coupling, respectively. On the other hand,  the effects of the transverse 
field $\Gamma$ are better shown in a phase diagram $T$ {\it versus} $\Gamma$ for 
constant values of $g$ and $J_{0}$. We also obtain a phase diagram $T$ {\it versus} 
$g$ in which $\Gamma$ and $J_{0}$ are related with $g$ based on the same arguments 
proposed in Ref. \cite{Magal05}. 
This procedure mixes the effects of both parameters in 
the the phase diagram. In the last section, we make 
our conclusions.

\section{General Formulation}

The model studied here is composed by interlattice Gaussian random spin-spin interaction \cite{KS},    
a intrasite local BCS pairing interaction (which favors double occupation of sites in each sublattice) with 
a transverse magnetic field applied $\Gamma$. Therefore, the hamiltonian is given by: 
\begin{eqnarray}
H= -\sum_{i_{a}j_{b}} J_{i_{a}j_{b}} S_{i_{a}}^{z} S_{j_{b}}^{z} 
- 2\Gamma\sum_{p}\sum_{i_{p}=1}^{N} S_{i_{p}}^{x}
\nonumber\\-
\frac{g}{N}
\sum_{p}\sum_{i_{p}j_{p}} c^{\dagger}_{i_{p}\uparrow} c^{\dagger}_{i_{p}\downarrow} c_{j_{p}\downarrow} c_{j_{p}\uparrow} 
\label{ham}
\end{eqnarray}
where the sums over $i_p~ (j_p)$ are run over the $N$ sites of each sublattice $p$ ($p=a$ or $b$).
The exchange interaction $J_{i_a j_b}$ is an independent random variable with the following Gaussian 
distribution:
\begin{equation}
P\left(J_{i_{a}j_{b}}\right) = \sqrt{\frac{N}{64\pi J^{2}}}  
\exp \left[- \frac {\left(J_{i_{a}j_{b}} + \frac{4}{N} J_{0} \right)^2}
{\frac{64J^{2}}{N}}\right].
\label{eq2}
\end{equation}

The spin operators in Eq. (\ref{ham}) are defined in terms of fermion operators:
\begin{equation}
S^{z}_{i_{p}} = \frac{1}{2}\left[\hat{n}_{i_{p}\uparrow} - \hat{n}_{i_{p}\downarrow}\right]
,~~S^{x}_{i_{p}} = \frac{1}{2}\left[c^{\dagger}_{i_{p}\uparrow}
c_{i_{p}\downarrow} +
c^{\dagger}_{i_{p}\uparrow} c_{i_{p}\downarrow}\right]
\label{eq4}
\end{equation}
where $\hat{n}_{i_p\sigma}$ gives the number of fermions at site $i_p$ with spin projection 
$\sigma= \uparrow$ or $\downarrow$. $c^{\dagger}_{i_p\sigma}$ and $c_{i_p\sigma}$
are the fermion creation and annihilation operators, respectively.

The problem is formulated in a path integral formalism in which the spin operators are represented
as anticommuting Grassmann fields ($\phi^{*},~\phi$). Therefore, the Grand canonical 
partition function is given by: 
\begin{equation}
Z= \int D[\phi^{*}\phi] \exp[~A~]
\end{equation}
with the action
\begin{equation}
\begin{split}
A=\int_{0}^{\beta}d\tau
\{\sum_{p,\sigma}
\sum_{i_p}[\phi^{*}_{i_p\sigma}(\tau)(
\frac{\partial}{\partial \tau}-\mu)\phi_{i_p\sigma}(\tau)]
\left.-H(\phi^{*}(\tau),\phi(\tau))\right\},
\label{eq5}
\end{split} 
\end{equation}
$\beta=1/T$ ($T$ is the temperature), $\tau$ is a complex time and $\mu$ is the chemical potential.
The Fourier decomposition of the time-dependent quantities is employed in Eq. (\ref{eq5}). 
The action can be write as $A=A_{\Gamma}+A_{SG}+A_{BCS}$ with:
\begin{equation}
A_{\Gamma}=\sum_{p}\sum_{i_{p}}\sum_{w}
\underline{\phi}^{\dagger}_{i_p}(\omega)[i\omega+\beta \mu+\beta\Gamma\underline{\sigma}^{x}]
\underline{\phi}_{i_p}(\omega),
\label{eqagamma}
\end{equation}
\begin{equation}
A_{SG}=\sum_{i_{a}j_{b}}\sum_{\omega^{'}}\beta J_{i_aj_b}S_{i_a}^{z}(\omega^{'})
S_{j_b}^{z}(-\omega^{'}),
\label{SG}
\end{equation}
\begin{equation}
S_{i_p}^z(\omega^{'})=\frac{1}{2}\sum_{\omega}\underline{\phi}^{\dagger}_{i_p}
(\omega+\omega^{'})
\underline{\sigma}^{z}\underline{\phi}_{i_p}(\omega),
\label{Si}
\end{equation}
\begin{equation}
A_{BCS}=\frac{\beta g}{N}\sum_{p}\sum_{i_pj_p}\sum_{\omega^{'}}
\rho^{*}_{i_p}(\omega^{'})\rho_{i_p}(\omega^{'})
\label{BCS}
\end{equation}
where $\rho_{i_p}(\omega^{'})=\sum_{w}\phi_{i_p\downarrow}(-\omega)\phi_{i_p\uparrow}(
\omega^{'}+\omega)$, $\underline{\sigma}^{\upsilon}~(\upsilon=x,~y$ or $z)$ denotes 
the Pauli matrices, $\underline{\phi}_{i_p}^{\dagger}=(\phi_{i_p\uparrow}^{*}(\omega)~~
\phi_{i_p\downarrow}^{*}(\omega))$ is a Grassmann spinor, and 
$\omega=(2m+1)\pi$ and $\omega^{'}=m\pi$ ($m= 0, \pm 1, \cdots$) are the Matsubara's frequencies.

The grand canonical potential is obtained within the static approximation which 
considers $\omega^{'}=0$ in Eqs. (\ref{SG})-(\ref{BCS})
\cite{Magal05,Zimmer06}.
The configurational averaged thermodynamic potential per site is obtained with the use of 
the replica method: 
$\beta \Omega
=-\frac{1}{2N}\displaystyle\lim_{n\rightarrow 0}(\langle Z^n \rangle_{J_{ij}} - 1)/n$
where the replicated partition function 
is:   
\begin{equation}
\langle Z^n \rangle_{J_{ij}}=\int \prod_{\alpha}D(\phi^{*\alpha}\phi^{\alpha})
\exp[A_{\Gamma}^{\alpha}+ A_{SG}^{st}+A_{BCS}^{st}]
\label{eq7}
\end{equation}
where $A_{\Gamma}^{\alpha}$ is given by Eq. (\ref{eqagamma}) with a sub-index $\alpha$,
\begin{equation}
A_{SG}^{st}=\sum_{i_{a}j_{b}}[\frac{8\beta^2 J^2}{N}(\sum_{\alpha}S_{i_a}^{\alpha}
S_{j_b}^{\alpha})^2
-\frac{2\beta J_{0}}{N}\sum_{\alpha}S_{i_a}^{\alpha}S_{j_b}^{\alpha}]
\label{asg}
\end{equation}

\begin{equation}
A_{BCS}^{st}=\frac{\beta g}{4N}\sum_{\alpha,p}\sum_{\upsilon=x,y}
[\sum_{i_{p}}\sum_{w}
\underline{\phi}^{'\alpha\dagger}_{i_p}(\omega)\underline{\sigma}^{\upsilon}
\underline{\phi}^{'\alpha}_{i_p}(\omega)]^2
\label{abcs}
\end{equation}
with the replica index $\alpha$ running from $1$ to $n$. In Eq. (\ref{abcs}), it is 
introduced the Nambu matrices $\underline{\phi}_{j}^{'\alpha\dagger}(\omega)$
and $\underline{\phi}_{j}^{'\alpha}(\omega)$
in which $\underline{\phi}_{j}^{'\alpha\dag}(\omega)=(\phi_{j\uparrow}^{*\alpha}(\omega)$ 
 $\phi_{j\downarrow}^{\alpha}(-\omega))$.

Eq. (\ref{asg}) can be rearranged reviewing the sums over different sublattices 
by square sums over the same sublattice. 
The replicated partition function is then linearized by using Hubbard-Stratonovich 
transformations. It inserts the replica-dependent auxiliary fields $q_{p}^{\alpha\beta}$,  
$m_{p}^{\alpha}$, $\eta_{R,p}^{\alpha}$ and $\eta_{I,p}^{\alpha}$ in Eq. (\ref{eq7}). 
The Gaussian integrals over these fields have been exactly performed in the 
thermodynamic limit by the steepest descent method. Therefore, the Grand canonical potential
is: 
\begin{eqnarray}
{\cal Z}(n)/N=
\beta g\sum_{\alpha,p}|\eta^{\alpha}_p|^{2}
-\beta J_0\sum_{\alpha}m^{\alpha}_a m^{\alpha}_b
+
\beta^{2}J^{2}
\sum_{\alpha\beta}q^{\alpha\beta}_a q^{\alpha\beta}_b-
\ln\Lambda_{\alpha\beta}^a\Lambda_{\alpha\beta}^b
\label{eq9}
\end{eqnarray}
where $\eta_{p}^{\alpha}=\eta_{R,p}^{\alpha}+i\eta_{I,p}^{\alpha}$ and
\begin{eqnarray}
&&\Lambda^{p}_{\alpha \beta}=\int \prod_{\alpha}D[\phi^{*\alpha}_{p}\phi_{p}^{\alpha}]
\exp[4\beta^2J^2\sum_{\alpha\beta} q_{p^{'}}^{\alpha\beta} S_{p}^{\alpha} S_{p}^{\beta}
\nonumber\\&& -~2\beta J_{0}\sum_{\alpha}m_{p^{'}}^{\alpha} S_{p}^{\alpha}
+A_{\Gamma,p}^{\alpha}+\beta g \sum_{\omega,\alpha}
\underline{\phi}_{p}^{'\dagger\alpha}\underline{\eta}_{p}^{\alpha} 
\underline{\phi}_{p}^{'\alpha}]
\label{eq10}
\end{eqnarray}
with the matrix $\underline{\eta}_{p}^{\alpha}=\eta_{R,p}^{\alpha}\underline{\sigma}^{x}+\eta_{I,p}^{\alpha}
\underline{\sigma}^{y}$. The fields $q_{p}^{\alpha\beta}$,  
$m_{p}^{\alpha}$ and $|\eta_{p}^{\alpha}|$ in Eq. (\ref{eq9}) are given by saddle-point equations, 
in which $q_{p}^{\alpha\beta}$ is related with the spin glass order parameter, 
$m_{p}^{\alpha}$ is the magnetization of the sublattice $p$, and $|\eta_{p}^{\alpha}|$ is an 
order parameter that indicates long range order where there is double occupation of sites 
in sublattice $p$.

In the present work, it is assumed the replica symmetric ansatz, which considers
$q_{p}^{\alpha\beta}=q_{p}$ for all $\alpha\neq\beta$, 
$q_{p}^{\alpha\alpha}=\overline{q}_{p}=\bar{\chi}_{p}+q_{p}$, $m_{p}^{\alpha}=m_{p}$, and 
$\eta_{p}^{\alpha}=\eta_{p}$ for all $\alpha$. 
The physical quantity $\beta \bar{\chi}_p$ is the static susceptibility when $J_{0}=0$.
The sums over replica indices are performed. It produces quadratic terms in Eq. (\ref{eq10}) that 
are linearized introducing new auxiliary fields in Eq. (\ref{eq9}).
The resulting
functional Grassmann integral is an exponential that sums quadratic forms 
of spinors and Nambu matrices. In order to perform the integral over the Grassmann 
fields, it can be used a matrix that mixes elements of spinors and Nambu matrices, such as:
\begin{eqnarray}
\Lambda^{p}_{\alpha \beta}&=&\int Dz_{p}\left\{ \int D\xi_{p} {\cal I}(z_p,\xi_p)\right\}^n
\label{eq101}
\end{eqnarray}
where $Dx=dx \mbox{e}^{-x^2/2}/\sqrt{2\pi}$ ($x=\xi_p$ or $z_p$),
\begin{equation}
{\cal I}(z_p,\xi_p)=\int D[\phi^{*}_{p}\phi_{p}]\exp[\sum_{\omega}
\underline{\Phi}^{\dagger}_{p}(w) \underline{G}_{p}^{-1}(w) \underline{\Phi}_{p}(w)]
\label{eq100}
\end{equation}
with
\begin{equation} 
\underline{\Phi}^{\dagger}_{p}(w) = \left[\phi^{*}_{p\uparrow}(w)
~~~\phi^{\ast}_{p\downarrow}(w)~~~\phi_{\downarrow p}(-w)~~~\phi_{p\uparrow}(-w)\right]
\label{eq12},
\end{equation}
\begin{equation}
\underline{G}^{-1}_{p}(\omega)=\left(
\begin{tabular}{cccc}
$i\omega+\zeta^+$ & $\beta\Gamma$ & $\beta g\eta_{p}$ & $0$\\
\\$\beta\Gamma$ & $i\omega+\zeta^-$ & $0$ & $-\beta g\eta_{p}$\\
\\$\beta g\eta_{p}^{*}$ & $0$ & $i\omega-\zeta^+$ & $-\beta\Gamma$\\
\\$0$ & $-\beta g\eta_{p}^{*}$ & $-\beta\Gamma$ & $i\omega-\zeta^-$
\end{tabular}
\right)
\label{eq13}
\end{equation} 
and ${\zeta^{\pm}}=\beta\mu_p \pm \beta h_{p}$. The internal field $h_{p}=J(
\sqrt{2q_{p^{'}}}z_p+\sqrt{2\bar{\chi}_{p^{'}}}\xi_p)- J_0m_{p^{'}}$, which acts
on the sublattice $p$, depends on the order parameters of sublattice $p^{'}$ ($p\neq p$) 
\cite{Zimmer06}. 

The functional integral in Eq. (\ref{eq100}) and the sum over the Matsubara's 
frequencies can be performed: 
\begin{equation}
{\cal I}(z_p,\xi_p)=\cosh\beta\sqrt{\mu^2+g^2\eta_{p}^{2}}+\cosh\beta\sqrt{\Delta_{p}}
\label{eq14}
\end{equation}
with $\Delta_p=h_{p}^{2}+\Gamma^2$. This result and Eq. (\ref{eq9}) are used to
express the thermodynamic potential as:
\begin{eqnarray}
2\beta \Omega=
-\beta J_0 m_a m_b+\beta^{2}J^{2}(\bar{\chi}_a\bar{\chi}_b+\bar{\chi}_a q_b+\bar{\chi}_b q_a)
\nonumber\\
+~\beta g(\eta_a^2+\eta_b^2)
-\sum_{p=a,b}\int_{-\infty}^{\infty} Dz_p\ln K_{p}(z_p)+\ln 4
\label{eq21}
\end{eqnarray}
where
\begin{equation}
K_{p}(z_p)=\left(
\cosh \beta g\eta_p+
\int_{-\infty}^{\infty}
D\xi_p\cosh\beta\sqrt{\Delta_p}\right)
\end{equation}
with $\mu=0$ to ensure the half-filling situation. The order parameters are given 
by the extreme condition of Eq. (\ref{eq21}):
\begin{equation}
m_p=
\int_{-\infty}^{\infty}Dz_p
\frac{
\int_{-\infty}^{\infty}D\xi_p\ \frac{h_p}{\sqrt{\Delta_p}}
\sinh\beta\sqrt{\Delta_p}}
{K_{p}(z_p)}
\label{eq22}
\end{equation}

\begin{equation}
\eta_p=\frac{1}{2}\int_{-\infty}^{\infty}Dz_p \frac{\sinh(\beta g \eta_p)}
{K_{p}(z_p)}
\label{eq23}
\end{equation}

\begin{equation}
q_p=
\int_{-\infty}^{\infty}Dz_p
\left[ \frac{
\int_{-\infty}^{\infty}D\xi_p 
\frac{h_p}{\sqrt{\Delta_p}}\sinh\beta\sqrt{\Delta_p}}
{K_{p}(z_p)}\right]^{2}
\label{eq24}
\end{equation}

\begin{equation}
\bar{\chi}_p=\int_{-\infty}^{\infty}Dz_p
\frac{\int_{-\infty}^{\infty}D\xi_p\frac{1}{\beta^2}\frac{\partial^2}{\partial h_p^2}
\cosh\beta\sqrt{\Delta_p}}{K_{p}(z_p)} -q_p
\label{eq25}.
\end{equation}

The stability of replica symmetric solution is analysed by Almeida-Thouless eigenvalue $\lambda_{AT}$:
\begin{equation}
\lambda_{AT}=1-2(\beta J)^4\prod_{p}\int_{-\infty}^{\infty}Dz_{p} \left(
\frac{I_p(z_p)}{(K_{p}(z_p))^2}\right)^2
\label{AT1}
\end{equation}
where
\begin{eqnarray}
I_p(z_p)=K_{p}(z_p)\int_{-\infty}^{\infty}D\xi_p\frac{1}{\beta^2}\frac{\partial^2}{\partial h_p^2}
\cosh\beta\sqrt{\Delta_p}
\nonumber \\ 
-~(\int_{-\infty}^{\infty}D\xi_p 
\frac{h_p}{\sqrt{\Delta_p}}\sinh\beta\sqrt{\Delta_p})^2.
\label{AT2}
\end{eqnarray}

\begin{figure}[t]
\begin{center}
\resizebox{.7\textwidth}{!}{
\includegraphics[angle=270]{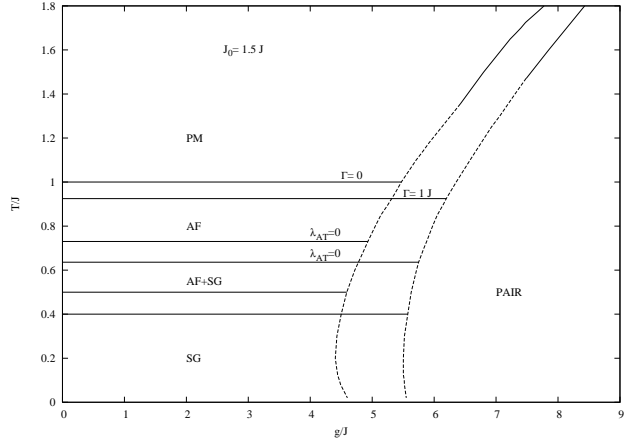}}
\caption{Phase diagrams as a function of $T/J$ and pairing coupling $g/J$ for $J_0=1.5 J$ and for two values of $\Gamma /J$: $\Gamma=0$ and
$\Gamma=1 J$. The full lines indicate second-order transitions while the dashed lines indicate first-order transitions. }  
\label{fig1}
\end{center}
\end{figure}

\section{Phase diagrams }

Numerical investigations of the order parameter equations (\ref{eq22}-\ref{eq25}) allow us 
to find three kinds of solutions. The SG solution corresponds to 
$q_{a}=q_{b}\neq 0$ (with $m_{a}=m_{b}=0$, $\eta_{a}=\eta_{b}=0$) while the AF solution 
is $m_{a}=-m_{b}\neq 0$ (with 
$q_{a}=q_{b}\neq 0$, $\eta_{a}=\eta_{b}=0$). The spin pairing solution (PAIR phase) 
corresponds to $\eta_{a}$ and $\eta_{b}$ 
different from zero while the rest of order parameters is zero. The instability of the 
replica symmetry (RS) solution of 
the SG is also investigated which allows us to identify the presence of a mixed phase AF+SG. 
This mixed phase corresponds to a replica symmetry breaking (RSB) 
SG with $m_{p}\neq 0$ ($p=a,b$) \cite{Magal05}. The emergence of each type of 
solution depends on the relationship among parameters $g$, $(J_{0})^{-1}$ (the degree of frustration) 
and $\Gamma$  given in units of $J$.

Therefore, we can build, in the beginning, two kinds of phase diagrams $T$ ($T$ is the temperature) 
{\it versus}: 
(a) $g$ ($g$ is the strength of intrasite pairing interaction) with $J_{0}$ and $\Gamma$ kept independents; (b) $\Gamma$ with $g$ and $J_{0}$ kept independents.
The first phase diagram can show directly the competition 
 among AF, SG and PAIR phases. 
The second one gives more precise information about the role of the transverse field on the transition lines present in the problem.  

\begin{figure}[t]
\begin{center}
\resizebox{.7\textwidth}{!}{
\includegraphics[angle=270]{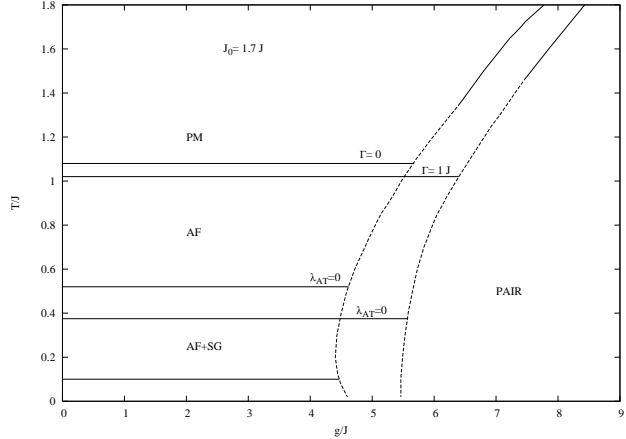}}
\caption{Phase diagrams as a function of $T/J$ and pairing coupling $g/J$ for $J_0=1.7 J$ and for two values of $\Gamma /J$: $\Gamma=0$ and
$\Gamma=1 J$. It is used the same convention as figure 1 for the transition lines.}  
\label{fig2}
\end{center}
\end{figure}

In Fig. 1, we show  the results for $T$ {\it versus} $g$ for
 $J_{0}=1.5J$ with $\Gamma=0$ and $J$. 
Therefore, in the region of small $g$ in  Fig. 1,  
it is quite clear that the parameter $J_{0}$ 
is related with 
the presence 
of magnetic solutions. Firstly, the AF solution 
appears below 
$T_{N}$. Then, 
when temperature is decreased, 
there is the onset of 
a mixed phase AF+SG at $T_{f}$. Finally, at lower temperature, there is a transition from 
AF+SG to a SG phase 
at $T_{g}$.     
The three magnetic transition lines mentioned above are second order. 
For large $g$, the solutions 
found for the order parameters indicate the existence of 
the PAIR phase \cite{magal99,magal00,Magal05} in which there is 
pairing formation in both sublattices. The role of the transverse field $\Gamma$ is 
also
clear in this particular phase diagram. The field decreases simultaneously
the magnetic transition temperatures $T_{N}$, $T_{f}$ and $T_{g}$. 
It also displaces the PAIR transition line $T_{1}(g)$  
in the sense that it is necessary to increase the parameter 
$g$ to find a PAIR solution in the problem. 

Fig. 2 shows the previous phase diagrams when 
the degree of frustration is 
 decreased ($J_{0}=1.7J$)   
with the  
transverse field kept $\Gamma=0$ and $J$. We can compare the results 
in the small and large $g$ region  in this figure
with the phase diagram  given in Fig. 1. The conclusion is direct, in the small $g$ region,
there are competing effects due to $J_{0}$ and $\Gamma$ \cite{Zimmer06}. The decrease of 
the degree of frustration enhances $T_{N}$. However, it decreases $T_{f}$ and $T_{g}$ as well. 
 On the other hand, the
whole set of magnetic transition temperatures $T_N, ~T_f$ and $T_g$ decreases with $\Gamma$. 
The superposition 
of both effects is responsible by   
the suppression of the SG phase in Fig. 2. 
The line transition $T_{1}(g)$ is not affected by the change 
of the degree of frustration $(J_{0})^{-1}$ while the transverse field $\Gamma$ 
has the same role 
as before, 
it displaces $T_{1}(g)$. 
 
\begin{figure}[t]
\begin{center}
\resizebox{.7\textwidth}{!}{
\includegraphics[angle=270]{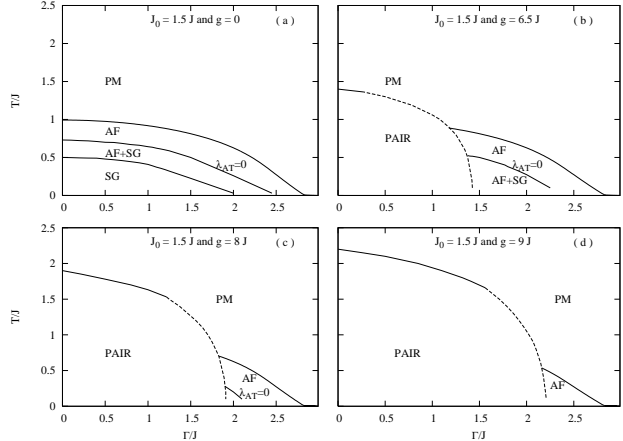}}
\caption{Phase diagrams as a function of $T/J$ and $\Gamma /J$ for $J_0=1.5 J$ and several fixed values of $g/J$: (a) $g=0$, (b) $g=6.5 J$,
(c) $g=8.0J$ and (d) $g=9.0J$. It is used the same convention as figure 1 for the transition lines. }  
\label{fig3}
\end{center}
\end{figure}

The numerical analyses indicate that the PAIR transition line $T_{1}(g)$  
is more complicated than the magnetic ones. It is a second order phase transition at higher temperatures and a 
first one at lower  temperatures, where there are multiple PAIR solutions. 
In this case, the stable solution minimazes the thermodynamic potential.
This same criterion is used to obtain the first-order boundary.
The transition lines can be also analysed by performing a Landau 
expansion of the thermodynamic potential in powers of the order parameters ($q_a$, $q_b$, 
$m_a$, $m_b$, $\eta_a$ and $\eta_b$). 
We can explore the symmetry of the parameters:  $q=q_a=q_b$, 
$\bar{\chi}=\bar{\chi}_a=\bar{\chi}_b$, $\eta=\eta_a=\eta_b$ and $m_a=-m_b$. 
Equation (\ref{eq21}) is expanded in powers of $q$, $\eta$ and $l=(m_a-m_b)/2$ ($l$ is the antiferromagnetic
order parameter), while $\bar{\chi}(q, l , \eta)$ is given by the saddle-point equation 
(\ref{eq25}). After some lengthy calculations, the Landau expansion of the thermodynamic 
potential is:
\begin{eqnarray}
2\beta\Omega=\beta^2 J^2\bar{\chi}_0-2\ln K_0 + 
A_2 l^2+
\nonumber \\
+B_2 q^2 +C_2 \eta^2 +C_4 \eta^4
\end{eqnarray}
with
\begin{eqnarray}
A_2=4\beta J_0(1-\beta J_0\bar{\chi}_0)l^2
\\
B_2=-\frac{\beta^2J^2}{2!}+ \beta^4J^4 \bar{\chi}^2_0,
\\
C_2=\beta g - \frac{\beta^2g^2}{2 K_0},
\label{B2}
\\
C_4=\frac{\beta^4 g^4}{4!K_0^2}(\frac{3 J^2}{g^2}K_0\bar{\chi}_0\bar{\chi}_2+3-K_0)
\label{B4}
\end{eqnarray}
where $K_0=1+\int_{-\infty}^{\infty}D\xi\cosh\sqrt{\Delta_0}~$,
\begin{equation}
\bar{\chi}_0=\frac{1}{K_0}\int_{-\infty}^{\infty}D\xi \xi^2
\frac{\sinh{\sqrt{\Delta_0}}}{\sqrt{\Delta_0}},
\end{equation}
\begin{equation}
\bar{\chi}_2=\frac{-\beta^2 g^2\bar{\chi}_0/K_0}{1+\beta^2 J^2[\bar{\chi}_0^2-
\int D\xi\xi^4(\frac{\cosh{\sqrt{\Delta_0}}}{\Delta_0}-\frac{\sinh{\sqrt{\Delta_0}}}{\Delta_0})/K_0]}
\end{equation}
and $\Delta_0=2\beta^2J^2\bar{\chi}_0\xi^2+\beta^2\Gamma^2$.
The tricritical point $T_{trict}$ can be obtained 
from  Eqs. (\ref{B2}-\ref{B4}) which show
that the transverse field $\Gamma$ affects the location of the $T_{trict}$. Actually, it
moves upwards $T_{trict}$ 
(see Figs. 1 and 2) \cite{magal00,Magal05}.

In Fig. 3, we show the phase diagram $T$ {\it versus} $\Gamma$ for $J_{0}=1.5J$ and 
$g=0,~6.5J,~8J$ and $9J$. The case in which there is no pairing coupling is shown in Fig. 3-a. 
In fact, this situation has been studied in Ref. \cite{Zimmer06} where the increase of 
$\Gamma$  leads the transition temperatures 
$T_{N}$, $T_{f}$ and $T_{g}$ towards their respective QCP's. In particular, the critical transverse field for AF transition 
can be obtained analytically
by expanding the sublattice magnetization $m_{p}$ (see Ref. \cite{Zimmer06}). The ordering   
$T_{g}<T_{f}<T_{N}$ is kept when $\Gamma$ increases, which is the reason why
the transition temperatures $T_{N}$, $T_{f}$ and $T_{g}$ are simultaneously depressed 
in Figs. 1 and 2. 
The increase of $g$  (see Figs. 1b, 1c, 1d) allows the existence of a PAIR solution 
which depends on, as discussed in Ref. \cite{Magal05}, of the ratio $\Gamma/g$  
as well as
the existence of a magnetic solution depends on $\Gamma/J_{0}$ 
in the present work. 
From this view point, 
the role of $\Gamma$ in Figs. 1 and 2 is clearly confirmed in Fig. 3, it tends 
to suppress any phase which is 
appearing as solution in the problem. Actually, the displacement of the PAIR phase 
in Figs. 1 and 2, when $\Gamma$ increases, reflects this effect. Moreover, the increase 
of $g$ also 
leads (if $J_{0}$ is kept constant) 
the PAIR phase 
to become
dominant.
It also moves upward the tricritical 
point
$T_{trict}$ as 
in Ref. \cite{Magal05}. 

\begin{figure}[t]
\begin{center}
\resizebox{.7\textwidth}{!}{
\includegraphics[angle=270]{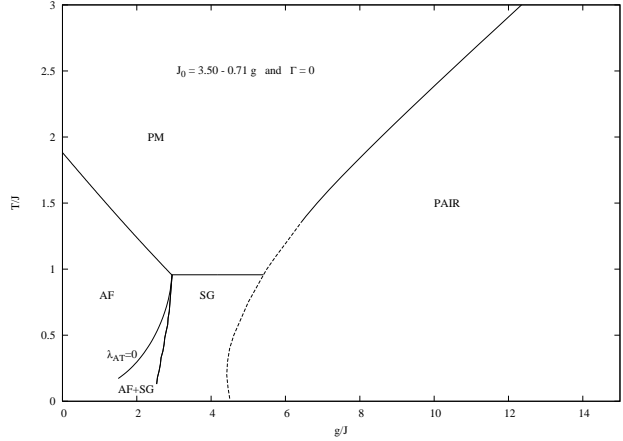}}
\caption{Phase diagram $T/J$ versus $g/J$ builds  for the relation 
$J_0=3.50-0.71g$ and $\Gamma=0$. The dashed line indicates a
first-order phase transition while the full lines indicate second-order phase transitions.} 
\label{fig4}
\end{center}
\end{figure}

The information contained in Figs. 1-3 can be displayed in a 
more adequate format
if we assume a relationship 
among the parameters $\Gamma$, $J_{0}$ and $g$. This kind of procedure has already been adopted 
in Ref. \cite{Magal05},
which is ultimately justified
by the fact that 
both RKKY and the pairing interaction in Eq. (\ref{ham}) are originated from the same source 
\cite{magal99}.  In the present case, $\Gamma$ would have the equivalent role of spin flipping part of the Heisenberg model \cite{Magal05}.
Besides, it is a more convenient 
format to compare with experimental results. 
Therefore, we assume the following relationship:
\begin{equation}
\Gamma=\alpha_{1} g+ \delta_{1}
\label{rel1}
\end{equation}
\begin{equation}
J_{0}=\alpha_{2} g+ \delta_{2}.
\label{rel2}
\end{equation}

The complicated interplay between $J_{0}$ and $\Gamma$ can be adjusted by the 
factors $\alpha_{1}$, $\alpha_{2}$, $\delta_{1}$ and $\delta_{2}$ in Eqs. (\ref{rel1})-(\ref{rel2}). We 
choose $\alpha_{2}<0$, which means that the degree 
of frustration 
 enhances with the increases of
the 
pairing coupling $g$. The factor $\delta_{2}$ is adjusted 
to guarantee the AF coupling 
and, with the remaining 
factors, also to maintain the transitions and the tricritical 
point located at the same scale as Figs. 1-3. 

\begin{figure}[t]
\begin{center}
\resizebox{.7\textwidth}{!}{
\includegraphics[angle=270]{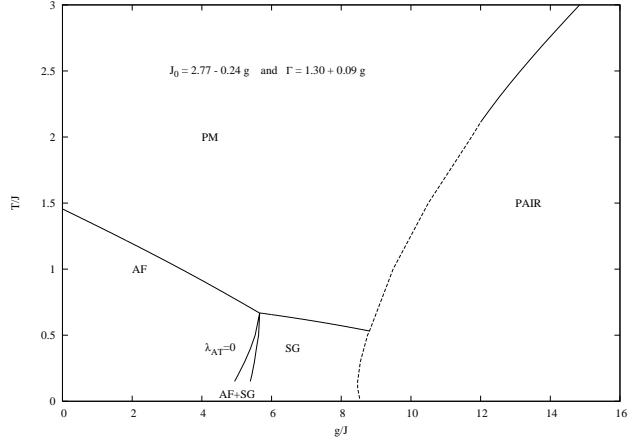}}
\caption{Phase diagram $T/J$ versus $g/J$ obtained from a relationship 
among $J_0/J$, $\Gamma/J$ and $g/J$ ($J_0=2.77-0.24g$ and $\Gamma=1.30+0.09g$).
It is used the same convention as figure 1 for the transition lines. }  
\label{fig5}
\end{center}
\end{figure} 

Fig. 4 shows the phase diagram $T$ {\it versus} $g$ with $\Gamma=0$. 
For that case, the behaviour of transition temperatures is obtained from the solution of order parameters 
(Eqs. (\ref{eq23})-(\ref{eq25})) together 
with the AT line  (Eqs. (\ref{AT1})-(\ref{AT2})) 
using only 
Eq. (\ref{rel2}). 
For small $g$ (small degree of frustration), 
there 
is  a transition from paramagnetism (NP)  to AF phase. 
Consequently, 
the N\'eel temperature decreases when $g$ is increased. Then, in a interval 
of $g$, 
there is direct transition from 
NP to SG phase  (in this case $T_g=T_f$). For large 
$g$ the 
PAIR phase is completely dominant.          
At 
temperature 
$T<T_{g}$, 
the situation 
is richer when $g$ increases as 
consequence of the RS lack of stability. The solutions found are 
AF at small $g$ and SG at some interval of $g$ as before. However, in a very small range of  
$g$, a  
mixed phase  AF+SG intermediated between AF and the SG phases appears. Then, after 
the sequence of second order transitions AF-AF+SG-SG, there is a first order 
boundary between SG and the PAIR phase. The location of the tricritical point $T_{trict}$ is not 
changed as compared with phase diagrams shown in Figs. 1 and 2 when $\Gamma=0$.
 
Figs. 5-6 show  the solution for the order parameters  when the 
transverse field $\Gamma$ is also tunned by Eq. (\ref{rel1}). The numerical factors 
$\alpha_{1}$ and $\delta_{1}$ can be used to adjust the strength of 
the transverse field as $g$ increases. In Fig. 5, $\Gamma$  affects 
the transition line $T_{N}$ and particularly $T_{g}$, which 
moves downwards when $g$ increases. The sequence of phases at  lower temperature 
is preserved as in Fig. 4. 
However, $\Gamma$ is not strong enough to lead $T_{g}$ to a QCP. 
On the other hand, 
the behaviour of the 
PAIR phase boundary $T_{1}(g)$ is affected as in Figs. 1-3 in which   
$T_{trict}$ moves upward  when $\Gamma$ increases.  In particular, it is possible to find one metastable SG solution into the PAIR phase 
below $T_{g}$ which 
keeps going 
downward.
Fig. 6 displays the situation where $\alpha_{1}$ and $\delta_{1}$ are adjusted 
to enhance the strength of $\Gamma$ as compared  with Fig 5.  For that case, the spin 
flipping induced by $\Gamma$ is strong enough to lead $T_{g}$ to a QCP while the 
$T_{trict}$ is obtained at a larger value than before.  For both cases of Figs 5 and 6, 
the tunning of $\Gamma$ still preserves the sequence of phases AF-AF+SG-SG at low temperature 
 likewise the case $\Gamma=0$ for a certain range of $g$.

\begin{figure}[t]
\begin{center}
\resizebox{.7\textwidth}{!}{
\includegraphics[angle=270]{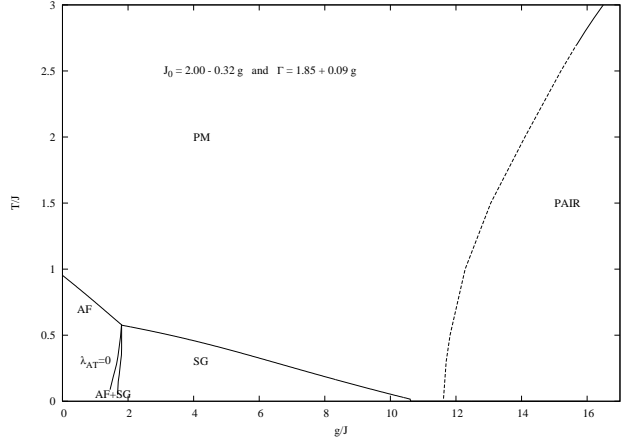}}
\caption{Phase diagram $T/J$ versus $g/J$ for the relations $J_0=2.00-0.322g$ and  $\Gamma=1.85+0.09g$. The same convention as figure 1
is used for the transition lines.}  
\label{fig6}
\end{center}
\end{figure}

\section{Conclusions} 

In the present paper, 
we have analysed the competition among 
antiferromagnetism (AF), spin glass (SG) and pairing formation phase (PAIR) in the presence of a 
quantum tunneling mechanism. 
The two-sublattice model used is composed by 
a Gaussian random interlattice Ising interaction (with mean $J_{0}$ and standard deviation $J$) \cite{KS}, 
an intralattice pairing interaction with an applied transverse field $\Gamma$. 
The 
partition function is calculated in the functional integral formalism 
in which the spin operators are given by bilinear combinations 
of Grassmann fields \cite{magal99,magal00,Magal05}. The saddle-point Grand Canonical potential is obtained    
within of static approximation (SA), the replica symmetry (RS) ansatz and in the half-filling. 
Particularly, the use of SA is justified  because 
our main interest is to study in detail the phase boundaries among AF, SG and PAIR 
phases when the spin flipping is activated by a transverse field $\Gamma$.

In the mean field theory presented, the phase transitions of the fermionic system 
defined in Eq. (\ref{ham})   
appear related  with pairing 
and  magnetic 
internal fields for each sublattice $p$. 
The magnetic one $h_{p}$ ($p=A$, $B$) 
has a random and AF components.  
In particular,  the AF part of $h_{p}$ depends 
on the sublattice magnetization $m_{p'}$ 
as well as
the random part is associated with the 
replica non-diagonal SG order parameter $q_{p'}$ and also
with $\overline{\chi}_{p'}=\overline{q}_{p'}-q_{p'}$ 
($\overline{q}_{p'}$    
is the replica diagonal SG order parameter), where $p\neq p'$ .  
In contrast, the pairing internal field applied in the sublattice $p$ 
depends on the PAIR order parameter $\eta_{p}$.
Furthermore, there is  
the presence of $\Gamma$ which tunes the spin flipping and, hence tends to suppress any kind 
of magnetic phase. 
The pairing formation is also affected, as it can be clearly seen in Fig. 3.

The solutions for  
$q_{p}$, $\overline{\chi}_{p}$, $m_{p}$ 
and  $\eta_{p}$ (PAIR order parameter) 
are located in a parameter space given (in units of $J$ ) by $J_{0}$,  $\Gamma$ and 
$g$ (the strength of the pairing interaction). 
Figures 1, 2 and 3 show the phase 
diagrams for several cuts in the previous space.
Thus, it is possible  
to identify how 
each parameter can favour one particular solution for a given temperature. 
Another important point is to locate the Almeida-Thouless line ($T_f$)
in such space. 
To take a typical case, in Fig.2, we present the phase diagram temperature {\it versus} $g$ for a 
$J_{0}=1.7J$ and $\Gamma=0$ and $J$. 
For lower $g$, the magnetic solutions are dominant with a sequence of second order phase transitions 
AF, a mixed phase AF+SG and SG at lower temperature. For larger $g$, 
the local pairing is dominant. 
Actually, the PAIR phase boundary $T_{1}(g)$ has a complex nature with the presence 
of a tricritical point $T_{trict}$ which is quite dependent on the transverse field $\Gamma$ 
(see Eqs. (\ref{B2}-\ref{B4})).       
 
In the phase diagrams 4, 5  and 6, we propose a relationship between the 
parameters $J_{0}$ and $\Gamma$ with $g$ (see Eqs. (\ref{rel1}-\ref{rel2})) based on the original derivation of the model 
given in Eq. (\ref{ham}) (see Ref.\cite{magal99}). 
This procedure allows to compare our results  with {the phase boundaries found in} experimental phase diagrams. 
For instance,  there are some 
similarities between the experimental situation for $Y_{1-x}$ $Ca_{x}Ba_{2}Cu_{3}O_{6}$ 
and the phase diagram shown in Fig. 5 as well as between  $U_{1-x}La_{x}Pd_{2}Al_{3}$ and the one shown in Fig. 6, 
if it is possible to associate the doping in those physical 
systems with the parameter $g$. 
For the first case, similarities such as  
the sequence of phases and, in particular, 
the presence 
of 
SG mixed with an AF background.
In terms of the present model, the decrease of experimental SG temperature transition  $T_{g}$ 
could be explained 
by the presence of quantum spin flipping mechanism which is not strong to lead $T_{g}$ towards 
a QCP.   
For the second one,
there are 
also 
similarities not only between the 
phase boundaries, 
but also with
the behaviour of $T_{g}$ which is depressed  to a QCP. 

To conclude, in this work we studied the thermodynamics of the model given 
in Eq. (\ref{ham}). Our goal is to obtain the corresponding phase boundaries and, then 
to mimic the global phase diagram of physical systems as  $Y_{1-x}$$Ca_{x}$$Ba_{2}$$Cu_{3}$$O_{6}$ and $U_{1-x}La_{x}Pd_{2}Al_{3}$.   
As last remarks
it should be noticed the role of 
$\overline{\chi}_{p}$, which carries the effects of disorder 
even at $T>T_{g}$, 
to determine 
the PAIR phase boundary   
$T_{1}(g)$. 
In the case $T>T_{g}$, 
$\overline{\chi}_{p}$ is equal to the  replica diagonal SG order parameter $\overline{q}_{p}$. 
In Fig. 6, 
because of the presence of QCP, this identity is true for the entire range of temperature after the QCP. 
It is well known that $\overline{q}_{p}$ can be 
written in terms of the
site occupation in the sublattices  \cite{NucB}. 
At the same time, the nature of $T_{1}(g)$ depends deeply on the transverse field $\Gamma$. 
This arises the question how the phase boundaries in the present problem would be affected 
 by the interplay between
the chemical potential $\mu$ and $\Gamma$ 
in situations like those shown in Figs. 5 and 6.    
It is also important to remark  that the precise location of the phase boundaries below $T_f$ needs 
RSB spin glass solutions, as for instance the boundary between SG and mixed phase and 
the first-order transition between SG (or mixed phase) and PAIR phase \cite{Oppermann1}. The study of the situation where $\mu\neq 0$ and 
the implementation of RSB 
will be object of a future work.

{\bf Acknowledgement:}
This work was partially supported by the Brazilian agencies CNPq (Conselho Nacional de Desenvolvimento Cient\'\i fico e Tecnol\'ogico) and CAPES (Coodena\c{c}\~ao de Aperfei\c{c}oamento de Pessoal de Nivel Superior).

\end{document}